\documentclass{elsart}

\usepackage{graphicx}

\begin{document}

\begin{frontmatter}

\title{The Stochastic Nature of Complexity Evolution in the
Fractional Systems}

\author{Aleksander~A.~Stanislavsky\corauthref{cor1}}
\corauth[cor1]{E-mail: alexstan@ri.kharkov.ua}

\address{Institute of Radio Astronomy, 4 Chervonopraporna
St., Kharkov 61002, Ukraine}

\begin{abstract}
The stochastic scenario of relaxation in the complex systems is
presented. It is based on a general probabilistic formalism of
limit theorems. The nonexponential relaxation is shown to result
from the asymptotic self-similar properties in the temporal
behavior of such systems. This model provides a rigorous
justification of the energy criterion introduced by Jonscher. The
meaning of the parameters into the empirical response functions is
clarified. This treatment sheds a fresh light on the nature of not
only the dielectric relaxation but also mechanical, luminescent
and radiochemical ones. In the case of the Cole-Cole response
there exists a direct link between the notation of the fractional
derivative (appearing in the fractional macroscopic equation often
proposed) and the model. But the macroscopic response equations,
relating to the Cole-Davidson and Havriliak-Negami relaxations,
have a more general integro-differential form in comparison with
the ordinary fractional one.


\end{abstract}


\end{frontmatter}

\section{Introduction}
Experimental investigations surely have established the relaxation
response of various complex systems (amorphous semiconductors and
insulators, polymers, molecular solid solutions, glasses, etc.) to
be non-exponential in nature \cite{1,2}. In particular, all types
of the empirical functions used to fit the dielectric data exhibit
the fractional-power dependence of the dielectric responses on
frequency and time. It is worth noticing that this unique property
is independent on any special details of examined systems. In the
past decades, a considerable attention has been paid to find a
theoretical explanation of the experimental results
\cite{2a,2b,3,4}. The main feature of all the dynamical processes
in the complex systems is their stochastic background. In this
framework one can expect that the macroscopic behavior of the
complex systems is governed by ``averaging principles'' like the
law of large numbers to be in force. However, to develop the
assumption is enough difficult. The point is that their
macroscopic evolution is not attributed to any particular object
taken from those forming the system. The problem of constructing
an "averaged" object representing the entire relaxing system is
not trivial. The description of the relationship between the local
random characteristics of complex systems and the universal
deterministic empirical laws is of great importance.

Many-body effects play a vital part in such systems. No wonder
that there exists a direct relationship, suggested in literature
\cite{2b,5,6,6aa}, between anomalous relaxation and anomalous
diffusion. One of more convenient languages for the description of
anomalous diffusion is the continuous random walk (CTRW) theory.
It occupies an important place for studying many physical
phenomena. The notation of CTRW was first proposed by Montroll and
Weiss in 1965 \cite{6a}. With their happy touch the CTRW
generalized a simple random walk. Although the term ``random
walk'' was introduced by Pierson in 1905, the formalism of simple
random walks was known else in the XVII-th century. The random
walk approach is based on the assumption that step changes are
made through equal time intervals. This was a first approximation
in various physical, chemical and economical models. For its turn
the CTRW theory allows a random waiting time among subsequent
random jumps. To sum up the long-term studies of this problem, the
recent, mathematically excellent works of Meerschaert, Scheffler
and Becker-Kern \cite{6b} have made its details ultimate clear. On
the one hand, this has allowed one to recognize the stochastic
processes responsible for the anomalous behavior. On the other
hand, the approach proposes a description of the anomalous
properties. In fact, it has established a close connection between
the stable distributions (from the theory of probability) and the
fractional calculus. This means that the nondifferentiable nature
of microscopic dynamics of components in the complex systems can
be transmitted to the macroscopic description of such systems in
the form of fractional operators \cite{6ba,6bb}. Consequently, the
CTRW method is very popular for physical applications connected
with anomalous diffusion, transport in disorder media, superslow
relaxation, etc. (see the perfect reviews \cite{6c,6d} and
references therein).

In this paper we suggest the probabilistic approach to the
analysis of evolution processes. Our approach is based on the
probabilistic formalism of limit theorems which provides tools to
relate the local random characteristics of complex systems to the
deterministic and universal relaxation laws regardless of the
specific nature of the systems considered. We attempt to answer
the following key questions related to the temporal evolution of
complex systems:
\begin{description}
\item[$\diamond$]what does mean a self-similarity in the evolution
of the complex stochastic systems (Section~2);\item[$\diamond$]
how to interpret theoretically the empirical deterministic
relaxation laws\\ (Section~3);\item[$\diamond$] what
characteristics of the internal structure of the complex systems
stand behind the empirical responses (Section~4);\item[$\diamond$]
what connection is between the micro/meso/macrocopic dynamics of
the relaxing systems and the macroscopic energy criterion
(Section~5);\item[$\diamond$] what role of the macroscopic
response equations is for the description of the relaxation
phenomena (Section~6).
\end{description}

Finally, we discuss some alternative models.

\section{Self-Similarity of Complex Systems}
The simplest traditional interpretation of relaxation phenomena is
based on the concept of a system of independent exponentially
relaxing objects (for example, dipoles) with different
(independent) relaxation rates \cite{6e}. Since any macroscopic
system consists of a finite number of objects, the approach gives
a discrete set of relaxation rates. Without a doubt, the
assumption may be valid for some relaxing systems. However, nether
finite nor infinite exponential series with different real rates
and contributions can result in the Havriliak-Negami (HN) and
Kohlrausch-Williams-Watts (KWW) response laws exactly. The only
opportunity to overcome this problem is a mathematical extension
of this model. Really, the transition from the discrete
``topology'' of the relaxation rate set to the continuous one
changes the situation in quality. By the integral transform
(resembling the Laplace transform and replacing the exponential
series) one can obtain the well-known empirical response
functions. In this case the temporary relaxation distribution
density is a continuous function and has no any narrow peaks that
could be interpreted as a manifestation of separate subsystems
(objects).

The relaxation, following the Cole-Cole (CC) law, may be developed
in a two-state system. Let $N$ be the common number of dipoles in
a dielectric system. If $N_\uparrow$ is the number of dipoles in
the state $\uparrow$, $N_\downarrow$ is the number of dipoles in
the state $\downarrow$ so that $N=N_\uparrow+N_\downarrow$. Assume
that for $t=0$ the system is stated in order so that the states
$\uparrow$ dominate, namely
\begin{displaymath}
\frac{N_\uparrow(t=0)}{N}=n_\uparrow(0)=1,\quad
\frac{N_\downarrow(t=0)}{N}=n_\downarrow(0)=0\,,
\end{displaymath}
where $n_\uparrow$ is the part of dipoles in the state $\uparrow$,
$n_\downarrow$ the part in the state $\downarrow$. Denote the
transition rates by $w$ defined from microscopic properties of the
system (for instance, according to the given Hamiltonian of
interaction and the Fermi's golden rule). In the case the kinetic
equation describing the ordinary relaxation (Debye law) takes the
form
\begin{equation}
\cases{\dot n_\uparrow(t)-w\,\{n_\downarrow(t)-
n_\uparrow(t)\}=0,&\cr \dot n_\downarrow(t)-w\,\{n_\uparrow(t)-
n_\downarrow(t)\}=0,&\cr}\label{eqa1}
\end{equation}
where, as usual, the dotted symbol  means the first-order
derivative. The steady state of the system corresponds to
equilibrium with $n_\uparrow(\infty)=n_\downarrow(\infty)=1/2$.
Clearly its response has an exponential character. However, this
happens to be the case for such dipoles that relax irrespective of
each other and of their environment. If the dipoles interact with
their environment, and the interaction is complex (or random),
their behavior already will not be exponential.

Assume that the interaction of dipoles with environment is taken
into account with the aid of the temporal subordination. Recall
that in the theory of anomalous diffusion the notation of
subordination occupies one of the most important places. So, a
subordinated process $Y(U(t))$ is obtained by randomizing the time
clock of a random process $Y(t)$ by means of a random process
$U(t)$ called the directing process.  The latter process is also
often referred to as the randomized time or operational time
\cite{7}. Generally speaking, the process $Y$ may be both random
and deterministic in nature. The anomalous diffusion theory
studies, as a rule, the subordination of random processes. We
intend to extend this approach to relaxation processes.

Let the time variable be a sum of random temporal intervals $T_i$
being nonnegative independent and identically distributed so that
the waiting times $T_i$ belong to an $\alpha$-stable distribution
($0<\alpha<1$). Then their sum
$n^{-1/\alpha}(T_1+T_2+\cdots+T_n),\, n\in\mathbf{N}$ converges in
distribution to a stable law with the same index $\alpha$
\cite{6b}. To determine a walker position at the true time $t$,
one needs to find the number of jumps up to time $t$. This
discrete counting process is $\{N_t\}_{t\geq
0}=\max\{n\in\mathbf{N}\mid \sum_{i=1}^nT_i\leq t\}$. Denote the
continuous limit of $\{N_t\}_{t\geq 0}$ by $S(t)$. For a fixed
time it represents the first passage of the stochastic time
evolution above that time level. The random process is
nondecreasing, and it can be chosen as a new time clock
(stochastic time arrow) \cite{7a}. The probability density of the
process $S(t)$ has the following Laplace image
\begin{equation}
p^{S}(t,\tau)=\frac{1}{2\pi j}\int_{Br} e^{ut-\tau
u^\alpha}\,u^{\alpha-1}\,du=t^{-\alpha}F_\alpha(\tau/t^\alpha)\,,
\label{eqb1}
\end{equation}
where $Br$ denotes the Bromwich path. This probability density has
a simple physical interpretation. It determines the probability to
be at the internal time (or so-called operational time) $\tau$ on
the real (physical) time $t$ \cite{7a}. The function $F_\alpha(z)$
can be expanded as a Taylor series. Besides, it has the Fox'
H-function representation
\begin{displaymath}
F_\alpha(z)=H^{10}_{11}\left(z\Bigg|{(1-\alpha,\alpha)\atop
(0,1)}\right)=\sum_{k=0}^\infty\frac{(-z)^k}{k!
\Gamma(1-\alpha(1+k))}\,,
\end{displaymath}
where $\Gamma(x)$ is the ordinary gamma function. In the theory of
anomalous diffusion the random process $S(t)$ is applied for the
subordination of L$\acute{\rm e}$vy (or Gaussian) random processes
\cite{6b,7b}. The inverse Levy process $S(t)$ accounts for the
amount of time that a walker does not participate in the motion
process \cite{7c}. If the walker participates all time in the
motion process, the internal time and the physical (external) time
would coincide.

As was shown in \cite{7a}, the stochastic time arrow can be
applied to the general kinetic equation. Then the equation
describing a two-state system takes the following form
\begin{equation}
\cases{\tilde D^\alpha
n_\uparrow(t)-w\,\{n_\downarrow(t)-n_\uparrow(t)\}=0,&\cr \tilde
D^\alpha
n_\downarrow(t)-w\,\{n_\uparrow(t)-n_\downarrow(t)\}=0,&\cr}\qquad
0<\alpha\leq 1,\label{eqc1}
\end{equation}
where $\tilde D^\alpha$ is the $\alpha$-order fractional
derivative with respect to time. Here we use the Caputo derivative
\cite{7d,7e}, namely
\begin{displaymath}
\tilde D^\alpha
x(t)=\frac{1}{\Gamma(n-\alpha)}\int^t_0\frac{x^{(n)}(\tau)}
{(t-\tau)^{\alpha+1-n}}\,d\tau,\quad n-1<\alpha<n,
\end{displaymath}
where $x^{(n)}(t)=D^nx(t)$ means the $n$-derivative of $x(t)$. The
relaxation function for the two-state system is written as
\begin{displaymath}
\phi_{\rm
CC}(t)=1-2n_\uparrow(t)=2n_\downarrow(t)-1=E_\alpha(-2wt^\alpha),
\end{displaymath}
where $E_\alpha(z)=\sum_{n=0}^\infty z^n/\Gamma(1+n\alpha)$ is the
one-parameter Mittag-Leffler function \cite{7ea}. Feller
conjectured and Pollard proved in 1948 that the Mittag-Leffler
function $E_\alpha(-t)$ is completely monotonic for $t\geq 0$, if
$0<\alpha\leq1$. Moreover, $E_\alpha(-t)$ is an entire function of
order $1/\alpha$ for $\alpha>0$ \cite{7}. It should be pointed out
that the relaxation function under interest corresponds to the CC
law.

The analysis of this problem will not be complete, unless one
consider it on the other hand. It turns out that the same result
mentioned above can be also obtained by another way
\cite{4,7j,7k}. If the relaxation rate of the {\it i}\,-th dipole
is equal to the value $b$, then the probability, that this dipole
did not change its initial orientation prior to an instant $t$, is
\begin{equation}
\mathrm{Pr}\,(\theta_i\geq\mathit{t}
\mid\beta_i=\mathit{b})=\exp(-\mathit{bt})\qquad\mathrm{for}\
\mathit{t}\geq 0,\ \mathit{b}>0. \label{eq1}
\end{equation}
The random variable $\beta_i$ denotes the relaxation rate of the
{\it i}\,-th dipole and the variable $\theta_i$ , the time needed
for changing its initial orientation. Let $\{\beta_i\}$ and
$\{\theta_i\}$ form sequences of nonnegative independent
identically distributed random variables. Following \cite{8} and
the law of total probability, one define the relaxation function
$\phi_i(t)$ for {\it i}\,-th dipole as a probability:
\begin{equation}
\phi_i(t)=\mathrm{Pr}\,(\theta_i\geq\mathit{t})=
\int^\infty_0\exp(-\mathit{bt})\,d\mathit{H_{\beta_i}(b)}\,,
\label{eq2}
\end{equation}
where $H_{\beta_i}$ is the distribution function of each
relaxation rate $\beta_i$. The form of a suitable function
$H_{\beta_i}$ should be found. In the system consisting of a large
number $N$ of relaxing dipoles, the relaxation function $\phi(t)$
has to express in terms of the probability that the entire system
will be without changing its initial state until $t$:
\begin{equation}
\phi(t)=\lim_{N\to\infty}\mathrm{Pr}\,(\mathit{A_N}\min(\theta_1,
\ldots,\theta_N)\geq t), \label{eq3}
\end{equation}
where $A_N$ is a normalizing constant. Let us observe that the
expression (\ref{eq2}) is the Laplace transform of the
distribution function $H_{\beta_i}(b)$:
\begin{displaymath}
\mathrm{Pr}\,(\theta_i\geq\mathit{t})=
\mathcal{L}(\mathit{H_{\beta_i}};t).
\end{displaymath}
Because of $\theta_i$ being independent, we get
\begin{displaymath}
\mathrm{Pr}\,\left(\min(\theta_1,
\ldots,\theta_N)\geq\mathit{\frac{t}{A_N}}\right)
=\left(\mathrm{Pr}\,(\theta_i\geq\mathit{\frac{t}{A_N}})
\right)^N=\left(\mathcal{L}(\mathit{H_{\beta_i}};\frac{t}{A_N})\right)^N.
\end{displaymath}
When $N$ tends to infinity, the {\it N}\,-th power of the Laplace
transform of the non-degenerate distribution function
$H_{\beta_i}$ converges to a non-degenerate limiting transform, if
and only if $H_{\beta_i}$ belongs to the domain of attraction of
the L${\rm\acute{e}}$vy-stable law \cite{7,8,9}.

In fact, the above limiting form is only determined by the
behavior of the tail of $H_{\beta_i}(b)$ for large $b$, i.\ e.\ by
asymptotic properties of $H_{\beta_i}(b)$. The detailed knowledge
of its other properties is not necessary. It is enough that the
distribution function $H_{\beta_i}$ belongs to a domain of
attraction of the L${\rm\acute{e}}$vy-stable law with the index of
stability $\alpha$. On the other words \cite{7}, the necessary and
sufficient condition for any $x>0$ is
\begin{equation}
\lim_{b\to\infty}\frac{1-H_{\beta_i}(xb)}{1-H_{\beta_i}(b)}=
x^{-\alpha}. \label{eq4}
\end{equation}
This condition can be interpreted as a type of self-similarity.
Really, for any $x>0$ and for large $b$
\begin{equation}
\mathrm{Pr}\,(\beta_i>\mathit{xb})\approx
x^{-\alpha}\,\mathrm{Pr}\,(\beta_i>\mathit{b}). \label{eq5}
\end{equation}
It is that the self-similarity is suggested as a fundamental
feature of relaxation phenomena \cite{5,10,11}. It should be
stressed here that in the approach this conclusion arises from the
pure probabilistic analysis, independently of the physical details
of dipolar systems. Thus, it can be carried over to other similar
cases of complex systems.

The randomness of the relaxation rates $\beta_i$ (1$\leq i\leq N$)
is motivated by the fact that in the complex systems an object has
not the only equilibrium state, but their states form a whole set
of metastable substates. Their configuration changes in a very
complicated way during their evolution. Each of the objects is
locked into a substate, and the distribution of relaxation rates
rejects any deterministic behavior of an individual object in the
complex system. The total survival probability of the whole system
has formally the same form
\begin{equation}
\mathrm{Pr}\,(\theta_i\geq\mathit{t})=\langle\exp(-\beta_it)\rangle
=\int^\infty_0\exp(-\mathit{bt})\,d\mathit{H_\alpha(b)}
\label{eq6}
\end{equation}
like (\ref{eq2}). However, now the form of d.\,f. $H_\alpha$ is
strictly fixed (i.\ e.\ it adheres to the
L${\rm\acute{e}}$vy-stable law), and the information about the
distribution functions $H_{\beta_i}$ is concentrated in the index
$\alpha$ of the stable law.

\section{Probabilistic Interpretation of Empirical
Laws}\label{par3}

Since the relaxation rate $b$ cannot be negative, the
L${\rm\acute{e}}$vy-stable laws are completely asymmetric
(supported on the nonnegative half-line) with $0<\alpha<1$. In
this case the relaxation function (\ref{eq3}) with
$A_N=N^{1/\alpha}$ is well defined and takes the KWW form
\begin{equation}
\phi_{\rm KWW}
(t)=\lim_{N\to\infty}\left(\mathcal{L}(\mathit{H_{\beta_i};
\frac{t}{N^{{\rm 1}/\alpha}}})\right)^N=\exp(-\,(At)^\alpha)\,,
\label{eq7}
\end{equation}
where $A$ is a positive constant \cite{12,13}. If $\alpha=1$, the
relaxation function (\ref{eq7}) becomes $\phi_D(t)=
\exp(-\,t/\tau_D)$ (Debye form), where $\tau_D=A^{-1}$ is the D
relaxation time. Mathematically, this case corresponds to the
degenerate case in (\ref{eq6}). For any fixed (deterministic)
constant $A$ we obtain the only expression (\ref{eq7}). In
general, the feature of $A$ is not necessarily true. It is
therefore reasonable to ask what will be, if the constant becomes
random. We will get the other relaxation laws.

To find the ``scenario'' leading to the observable relaxation laws
different from the KWW form, let us note of the fact that the
relaxation function for the CC response can write in the form
\begin{equation}
\phi_{\rm CC}
(t)=\int^\infty_0\exp(-\,(t/\lambda)^a)\,dT_a(\lambda/\tau_{\rm
CC})= E_a(-(t/\tau_{\rm CC})^a)\,, \label{eq8}
\end{equation}
where $E_a(z)$ is the one-parameter Mittag-Leffler function,
$T_a(\lambda/\tau_{\rm CC})$ the one-sided
L${\rm\acute{e}}$vy-stable probability distribution with the index
$0<a\leq 1$, and $\tau_{\rm CC}$ is constant. It is useful to
recall that from the subordination approach the CC relaxation
response is expressed as
\begin{equation}
\phi_{\rm CC}(t)=\int^\infty_0\phi_{\rm
D}(\tau)\,p^S(t,\tau)\,d\tau=
\int^\infty_0\exp(-\mu\tau)\,p^S(t,\tau)\,d\tau= E_\alpha(-\mu
t^\alpha)\,.\label{eq8a}
\end{equation}
Here the parameter $\mu$ is constant, and the contribution of
irregular changing dipole orientations in the macroscopic
evolution of the system is derived from the probability density of
the directing process $S(t)$. The approach of Weron and Jurlewich
\cite{7j,7k} is based on the other conception. It is that each
individual dipole in a complex system relaxes exponentially, but
their relaxation rates are different and obey a probability
distribution (continuous function). However, the subordination
approach brings advantages in deriving a rather simple macroscopic
equation for the description of the CC relaxation response.

The result (\ref{eq8}) may be interpreted as a weighted average
(or as randomizing the parameter $\lambda$) of the stretched
exponential relaxation (\ref{eq7}) respect to the distribution
function $T_a(\lambda)$ of the scale parameter $\lambda$. This
idea works not only for the CC relaxation. Really, let $Q_a$ be
such a random value that its Laplace transform is the stretched
exponential function
\begin{equation}
\langle\mathrm{e}^{\mathit{-s\,Q_a}}\rangle=\int^\infty_0
\exp(-\mathit{st})\,h_a(t)\,dt\,,\qquad
\mathrm{0}<\mathit{a}\leq\mathrm{1}.\label{eq9}
\end{equation}
Then the random value $Q_a$ is distributed according to the
one-side L${\rm\acute{e}}$vy-stable law with the probability
distribution function $h_a(t)$ with $0<a<1$ (see details, for
example, \cite{13}). Now let the random value $G_b$ be independent
of $Q_a$ and distributed according to the gamma law \cite{15}
defined by the probability distribution function
\begin{displaymath}
g_b(t)=\frac{1}{\Gamma(b)}\,t^{b-1}{\mathrm{e}}^{-t},\qquad b>0,\
t>0.
\end{displaymath}
In this connection it should be pointed out that the Laplace
transform of $G_b$ takes the form
\begin{equation}
\langle\mathrm{e}^{\mathit{-s\,G_a}}\rangle=\int^\infty_0
\exp(-\mathit{st})\,g_b(t)\,dt=\frac{\mathrm{1}}{(\mathrm{1}+
\mathit{s})^b}\,.\label{eq10}
\end{equation}
For the random value $B\,Q_a\,(G_b)^{1/b}$ one obtains
\begin{eqnarray}
\langle\mathrm{e}^{\mathit{-s\,B\,Q_a\,G_a}}\rangle
&=&\left(\int^\infty_0\exp\,(-\mathit{Bst^{1/a}})\,h_a(s)
\,ds\right)g_b(t)\,dt\\ \nonumber &=&\int^\infty_0
\exp\,(-(sB)^at)\,g_b(t)\,dt=\frac{\mathrm{1}}{(\mathrm{1}+
(\mathit{Bs})^a)^b}\,,\label{eq11}
\end{eqnarray}
where the positive (arbitrary) constant $B$ is a scale parameter.
The frequency-domain response $\phi^\star(\omega)$ is related to
the relaxation function $\phi(t)$ by the one-sided Fourier
transform:
\begin{equation}
\phi^\star(\omega)=\int^\infty_0\mathrm{e}^{\mathit{i\omega t}}
\left(-\, \frac{\mathit d\phi(t)}{\mathit
dt}\right)\,\mathit{dt}.\label{eq12}
\end{equation}
As it is well known \cite{5}, the (dielectric) susceptibility
$\chi(\omega)$ is directly connected with $\phi^\star(\omega)$ by
the formula:
\begin{displaymath}
\phi^\star(\omega)=\frac{\chi(\omega)-\chi_\infty}{\chi_0-
\chi_\infty},
\end{displaymath}
where the constant $\chi_\infty$ represents the asymptotic value
of $\chi(\omega)$, and $\chi_0$ is the value of the opposite
limit. Clearly, the process $B\,Q_a\,(G_b)^{1/b}$ leads to the
following time-frequency response
\begin{equation}
\phi^\star_{\rm HN}(\omega)=\frac{\mathrm{1}}{(\mathrm{1}+
(\mathit{iB\omega})^a)^b}.\label{eq13}
\end{equation}
This is just the HN relaxation response. Evidently, for $b=1$ the
expression corresponds to the Cole-Davidson (CD) empirical law.
When the random value $(G_1)^{1/c}$ follows the Weibull
distribution \cite{15} with the shape parameter equal to $c$, we
arrive at the CC relaxation. Thus, the KWW, CC, CD and HN
relaxation functions are very close in connection from the
probabilistic point of view to the random processes associated
with the relaxation.

In this connection it should be pointed out that the evolution of
$n_\uparrow(t)$ and $n_\downarrow$ in Eq. (\ref{eqb1}) can be
connected with the Mittag-Leffler distribution. Let $Z_n$ denote
the sum of $n$ independent random values with the Mittag-Leffler
distribution. Then the Laplace transform of $n^{-1/\alpha}Z_n$ is
$(1+s^\alpha/n)^{-n}$, which tends to $e^{-s^\alpha}$ as $n$ tends
to infinity. Following Pillai \cite{15a}, this demonstrates an
infinity divisibility of the Mittag-Leffler distribution. By
virtue of the power asymptotic form (long tail) the distribution
with parameter $\alpha$ is attracted to the stable distribution
with exponent $\alpha$, $0<\alpha<1$. The property of the
Mittag-Leffler distribution allows one to develop a corresponding
stochastic process. The process (called Mittag-Leffler's) arises
of subordinating a stable process by a directing gamma process
\cite{15a}. In this case the relaxation function has the
Havriliak-Negami form
\begin{equation}
\phi_{\rm
HN}(t)=1-\sum_{k=0}^\infty\frac{(-1)^k\Gamma(b+k)}{k!\Gamma(b)\Gamma(1+ab+ak)}
\left(t/\tau_{\rm HN}\right)^{ab+ak}\,, \label{eq13a}
\end{equation}
where $a$, $b$, $\tau_{\rm HN}$ are constant. The one-side Fourier
transformation of the relaxation function gives
\begin{equation}
\phi^\star_{\rm HN}(\omega)=\int^\infty_0e^{-i\omega
t}\,\left(-\frac{d\phi_{\rm HN}(t)}{dt}\right)\,dt=
\frac{1}{(1+(i\omega\tau_{\rm HN})^a)^b}\,. \label{eqe13b}
\end{equation}
This result also corresponds to the well-know HN empirical law.
Thus, the HN relaxation can be explained from the subordination
approach, if the hitting time process of dipole orientations
transforms into the Mittag-Leffler process. For that the hitting
time process has an appropriate distribution attracted to the
stable distribution. The subordination of the latter results just
in the Mittag-Leffler process. It is interesting to observe that
the L\'evy process subordinated by another L\'evy one leads again
to the L\'evy process, but with other index \cite{15b}. Observe
that in this point the subordination approach is almost equivalent
to the approach studied in \cite{7j}.

\section{Internal Structure of Complex Systems}
In any dielectric (complex) system under an week external electric
field (external action) only a part (active dipoles or objects) of
the total number $N$ of dipoles is directly governed by changes of
the field. But even those dipoles, not contributing to the
relaxation dynamics, can have an effect on the behavior of active
dipoles. This means that the {\it i}\,-the active dipole interacts
with $N_i-1$ inactive neighbors forming a cluster of size $N_i$.
The number $K_N$ of active dipoles in the system is equal to the
number of clusters. The sum of the clusters exceeds $N$, the size
of the system. Because of the screening effects the active dipoles
can ``see'' only some of their active neighbors. If so, the
cooperative regions built upon the active dipoles will appear. The
number of the such mesoscopic regions is determined by their sizes
$M_1, M_2, \ldots$. The contribution of each region to the total
relaxation rate is a sum of the contributions of all active
dipoles over the region. Generally speaking, the sums are random.
Hence, the {\it j}\,-th region has its relaxation rate, say
$\overline{\beta_{jN}}$, equal to
\begin{displaymath}
\overline{\beta_{jN}}=\sum^{M_1+\ldots+M_j}_{i=M_1+
\ldots+M_{j-1}+1}\beta_{jN}.
\end{displaymath}
For $j=1$ the latter expression is simply the sum
$\overline{\beta_{1N}}=\sum^{M_1}_{i=1}\beta_{iN}$. Next for $j=2$
it takes the form
$\overline{\beta_{2N}}=\sum^{M_1+M_2}_{i=M_1+1}\beta_{iN}$ and so
on. The relaxation function of the whole system
\begin{displaymath}
\phi(t)=\langle{\mathrm{e}^{-\mathit t\tilde{\beta}_N}}\rangle
\end{displaymath}
is provided by the total relaxation rate $\tilde{\beta}_N$ as the
sum of the contributions over all cooperative regions:
\begin{displaymath}
\tilde{\beta}_N=\sum^{L_N}_{j=1}\overline{\beta_{jN}}.
\end{displaymath}
As a rule, the relaxing systems consist of a large number of
dipoles so that the limit transition
$\tilde{\beta}=\lim_{N\to\infty}\tilde{\beta}_N$ is valid (in
practice, $N\approx 10^5$ and more is enough). Limit theorems for
the random sums have been recently established in \cite{15c}.

The number of dipoles directly engaged in the relaxation process
is random as well as their locations are random too. Obviously,
all the quantities $N_i$'s, $M_j$'s, $\beta_{iN}$'s and those
defined by them, are random values. Their stochastic
characteristics determine the total relaxation rate
$\tilde{\beta}_N$, but they are not known. Nevertheless, on the
basis of the limit theorems of probability theory, the
distribution of the limit $\tilde{\beta}$ (for the large relaxing
systems) can be defined, even with rather information about the
distributions of micro/mesoscopic quantities.

In the approach it is quite enough to consider stochastically
independent sequences of random values $N_i$'s, $M_j$'s,
$\beta_{iN}$'s. Each sequence consists of independent and
identically distributed nonnegative random values that have either
finite expected value finite or long-tailed distribution. Then the
total relaxation rate $\tilde{\beta}$ takes the form corresponding
to one of the empirical responses (see Section~\ref{par3}). It
should be noted that the distribution of a nonnegative random
value, say $X$, has a long tail, if and only if the tail Pr($X>x$)
fulfills the condition
\begin{equation}
\lim_{x\to\infty}\frac{\mathrm{Pr}(\mathit{X>x})}{x^{-\gamma}}=
\mathrm{const}>\mathit{0}\label{eq14}
\end{equation}
for some $0<\gamma<1$ so that for value $x$ the tail exhibits the
fractional power law $x^{-\gamma}$ \cite{7,8}. Many different
continuous and discrete distributions are well known to satisfy
the condition (\ref{eq14}). Classical examples are completely
asymmetric L${\rm\acute{e}}$vy-stable laws as well as the Pareto
and Burr distributions with an appropriate choice of their
parameters \cite{7}. To obtain the discrete distributions with
long tails, one should apply a quantization procedure to the above
continuous examples \cite{8}. If the distribution of random value
$X$ has long tail, then the expected value $\langle X\rangle$ is
infinite. The finiteness of the expected value and long-tail
property (\ref{eq14}) can be presented only on different levels
(theirs are three: an active dipole $\rightarrow$ a cluster
$\Rightarrow$ a cooperative region) of the complex system. To sum
up, Table~\ref{tab1} shows the connection between the internal
properties of complex system's dynamics and the empirical
relaxation responses, as well as the physical sense of the
parameters characterizing the responses. The proposed approach
leads to a very general scenario of relaxation, from the
stochastic nature of microscopic dynamics through the hierarchical
structure of parallel multi-channel processes to the deterministic
macroscopic laws of relaxation given by (\ref{eq7}) and
(\ref{eq13}).
\begin{table}  \caption{The connection between the internal properties of
  complex systems and their relaxation response (the notations
  of the column ``Parameters'' correspond to (\ref{eq7}) and
  (\ref{eq13}); the constant $\gamma$ according to (\ref{eq14})).}\centering{
  \begin{tabular}{|c|c|c|c|c|}\hline
    &  &  &  &  \\
 Law & Parameters & $N_i$ & $M_j$ & $\beta_{iN}$  \\
    &   &  &  &  \\
 \hline
    &  &  &  &   \\
    & $a=1$ & & & \\
    D &$b = 1$ & $\langle N_i\rangle <\infty$ & $\langle M_j\rangle
    <\infty$& $\langle\beta_{iN}\rangle <\infty$ \\
    &  &  &  &  \\ \hline
    &  &  &  &  \\
    KWW & $0<\alpha<1$ & $\langle N_i\rangle <\infty$ &
    $\langle M_j\rangle <\infty$& long tail \\
    &  &  &  & $\gamma=\alpha$  \\
    &  &  &  &  \\  \hline
    &  &  &  &  \\
    & $a=1$ & & long tail & \\
    CD & $0<b<1$ & $\langle N_i\rangle <\infty$ & $\gamma=b$ &
    $\langle\beta_{iN}\rangle <\infty$ \\
    &  &  &  &  \\  \hline
    &  &  &  &  \\
    & $0<a<1$ &long tail & &long tail \\
    CC & $b=1$ & $\gamma=a$ & $\langle M_j\rangle <\infty$ &
    $\gamma=a$ \\
    &  &  &  &  \\ \hline
    &  &  &  &  \\
    & $0<a<1$ & long tail & long tail & long tail \\
    HN & $0<b<1$ & $\gamma=a$ & $\gamma=b$ &
    $\gamma=a$ \\
    &  &  &  &  \\  \hline
  \end{tabular}
\label{tab1}}
\end{table}

\section{Energy Criterion}
The common property of the empirical relaxation laws is that they
exhibit the high-frequency power law in the susceptibility:
\begin{displaymath}
\chi(\omega)\propto\left(\frac{\mathit{i}\omega}{\omega_{\mathit
p}}\right)\qquad \mathrm{for}\ \omega\gg\omega_{\mathit p},
\end{displaymath}
where the exponent $n$ falls in range (0,1) and the constant
$\omega_p$ is the loss peak frequency. As a consequence, for large
$\omega$ the ratio of the imaginary to real components of the
susceptibility $\chi(\omega)=\chi '(\omega)-i\chi ''(\omega)$
becomes a constant of degree $n$:
\begin{equation}
\frac{\chi '' (\omega)}{\chi
'(\omega)}=\cot\left(n\frac{\pi}{2}\right)\qquad\mathrm{for}\
\omega\gg\omega_{\mathit p}.\label{eq15}
\end{equation}
However, the D response has not the property. It should be noted
the physical significance of expression (\ref{eq15}). At high
frequencies the ration of the macroscopic energy lost per radian
to the energy stored at the peak is independent of frequency.

Jonscher \cite{2} has advanced a hypothesis that the fact is based
on the identical property of individual structural elements of the
systems. This explains the universality in the large scale
behavior of complex systems, but needs for the precise derivation.
In the framework of the proposed and mentioned-above model the
physical intuition can be strictly argumentative. Really, the
condition (\ref{eq14}) applied to any relaxation rate $\beta$
leads to the scaling property of the relaxation-rate distribution
at large $b$ (see also (\ref{eq5})). The asymptotic behavior of
the distribution is connected with the short-time asymptotic
properties of the associated relaxation function $\phi(t)$, and
the response function as its derivative $f(t)=-d\phi(t)/dt$ takes
the form
\begin{displaymath}
f(t)\propto t^{\gamma-1}\,U(t)
\end{displaymath}
for $t\to 0$, where $U(t)$ is a slowly varying function so that
$\lim_{t\to 0}U(ct)/U(t)=1$ for any constant $c>0$. It may be
easily verified that the short-time behavior of $f(t)$ corresponds
to the high-frequency properties of the susceptibility
$\chi(\omega)$:
\begin{displaymath}
\chi(\omega)=\chi'(\omega)-i\chi''(\omega)\propto
(i\omega)^{-\gamma}\,U(1/\omega).
\end{displaymath}
The result yields straightforwardly the energy criterion
(\ref{eq15}) with $n=1-\gamma$. The long-tail property of
micro/meso/macroscopic relaxation rate with the parameter $\gamma$
leads to micro/meso/macroscopic energy criterion with the
characteristic constant $1-\gamma$. The analysis of the model
shows \cite{7k} that in the HN, CC  and KWW responses the energy
criterion is the case for all micro/meso/macro levels, and the
constant $n$ for the HN case is defined not only the long-tail
property of the distribution of cluster sizes, but one of
cooperative-region sizes. In the CD case the microscopic energy
criterion is absent. The high-frequency power law of this response
results only from the long-tail property of the distribution of
cooperative-region sizes.

\section{Macroscopic Response Equations}
It is well known (see e.\ g.\ \cite{16}) that the relaxation
function $\phi(t)$ has to fulfil the two-state master equation
\begin{equation}
\frac{\mathit{d}\phi(t)}{\mathit{dt}}=-r(t)\,\phi(t),\qquad
\phi(0)=1,\label{eq24}
\end{equation}
where the nonnegative, time-dependent value $r(t)$ is the
transition rate of the relaxing system (i.\ e.\ the probability of
transition per unit time). This is a macroscopic deterministic
equation. Although the equation does not contain any (for example,
micro/mesoscopic and so on) details about relaxation processes, it
is convenient for practical purpose because of its simplicity. So,
in the case of the KWW relaxation $r(t)=a\,A^at^{a-1}$. For the D
relaxation the equation (\ref{eq24}) has the simplest form
$r(t)=r_0=$ const. Then, both relaxation function $\phi(t)$ and
response function $f(t)=-\, d\phi(t)/dt$ satisfy the same
equation, and their expressions coincide. However, this
equivalence is wrong for nonexponential relaxations. Following the
old notation (\cite{6e}, p.~8), the response function is a
pulse-response function of the polarization. Any relaxation
function or ``decay function of the polarization'' (by definition,
see \cite{6e}) tends to 1 for $t\to 0$. But some response
functions have a singularity in zero, as appears, for example,
from the CD response function. In general, the response function
$f(t)$ can be a solution of the first-order differential equation
with variable coefficients:
\begin{equation}
\left[\mathit r^{\mathrm 2}(t)-\frac{\mathit dr(t)}{\mathit
dt}\right]\,f(t)+r(t)\,\frac{\mathit f(t)}{\mathit
dt}=0.\label{eq25}
\end{equation}
When the relaxation obeys the CC, CD, HN laws, the equations
(\ref{eq24}) and (\ref{eq25}) are not just simple because of a
sufficiently complicated expression for $r(t)$. It may be
attempted to transform (\ref{eq24}) and (\ref{eq25}) in a
integro-differential form simpler than the input ones. Really, the
way gives some progress. The CC relaxation and response functions
can be expressed as a solution of the fractional differential
equation.

In this case we have
\begin{displaymath}
\phi_{\rm CC}(t)=E_a(-(t/\tau)^a),
\end{displaymath}
and the response function is written in the following series
representation
\begin{displaymath}
f_{\rm CC}(t)=\sum^\infty_{k=0}\frac{(-1)^k(t/\tau)^{a(1-k)-1}}
{\tau\Gamma [a(1+k)]}.
\end{displaymath}
According to the book of Miller and Ross \cite{17}, the
one-parameter Mittag-Leffler function $E_a(-(t/\tau)^a)$ satisfies
the identity
\begin{displaymath}
J^{1-a}[D\,E_a(-(t/\tau)^a)]=-\frac{1}{\tau^a}E_a(-(t/\tau)^a),
\end{displaymath}
where $D$ denotes the usual differentiation operator $d/dt$, and
the operator $J^{1-a}$ is the Riemann-Liouville fractional
integral having the form
\begin{equation}
J^{\nu}x (t)=\frac{1}{\Gamma(\nu)}\int^t_0(t-s)^{\nu-1}\,x(s)\,ds.
\label{eq26}
\end{equation}
From this it follows that for the CC relaxation the relaxation
function fulfils
\begin{equation}
J^{1-a}[D\,\phi_{\rm CC}(t)]=-\frac{1}{\tau^a}\,\phi_{\rm CC}(t),
\label{eq27}
\end{equation}
and the response function does
\begin{equation}
D[J^{1-a}f_{\rm CC}(t)]=-\frac{1}{\tau^a}\,f_{\rm CC}(t).
\label{eq28}
\end{equation}
In this connection it should be pointed out that equation
(\ref{eq27}) is expressed in terms of the fractional integral of
the ordinary derivative, and (\ref{eq28}) is in terms of the
ordinary derivative of the fractional integral. The equations
(\ref{eq27}) and (\ref{eq28}) are fully equivalent to (\ref{eq24})
and (\ref{eq25}) with the transition rate
\begin{displaymath}
r_{\rm CC}(t)=-[E_a(-(t/\tau)^a)]^{-1}D[E_a(-(t/\tau)^a)].
\end{displaymath}
In fact, this is only another formulation of (\ref{eq24}) and
(\ref{eq25}), in terms of the Green function (see more details,
for example, in \cite{18}).

With macroscopic equations for the CD and HN responses the
situation is more intricate. From the above consideration it is
seen that the responses are results of the (not simple) integral
transformations. This means that probably, the CD and HN responses
satisfy enough complicated macroscopic equations. Therefore, it
should be given consideration.

An interesting idea was suggested in \cite{3}. It proceeds from
the fact that the ordinary equation of exponential relaxation can
be written in the form
\begin{displaymath}
\exp(-\Omega_0t)\frac{d}{dt}\exp(\Omega_0t)f_{\rm D}(t)=0\,,
\end{displaymath}
where $\Omega_0$ is constant. By the direct substitution \cite{3}
it is easily verified that the CD response function is a solution
the following equation
\begin{equation}
\exp(-\Omega_0t)D^\nu[\exp(\Omega_0t)f_{\rm
CD}(t)]=0\,.\label{eq29}
\end{equation}
Here the fractional derivative
\begin{displaymath}
D^{\nu}[x(t)]=1/\Gamma(1-\nu)\frac{d}{dt}\int^t_0(t-s)^{-\nu}\,
x(\tau)\,d\tau
\end{displaymath}
is defined as well as in \cite{3}. Next, Eq. (\ref{eq29}) can be
again generalized, namely
\begin{equation}
\exp(-\Omega_0t^\mu)D^\nu[\exp(\Omega_0t^\mu)f_{\rm gen}
(t)]=0\,.\label{eq30}
\end{equation}
Now its solution
\begin{displaymath}
f_{\rm gen}(t)\sim t^{\nu-1}\exp(-\Omega_0t^\mu)\,,
\end{displaymath}
describes both CD ($\mu=1$ and $0<\nu<1$) and KWW ($0<\mu=\nu<1$)
response functions.

In \cite{3} it is proved that the macroscopic equation like
$(D^\varepsilon+\Omega)^{\alpha/\varepsilon}\,f(x)=0$ is well for
the HN response. Their conclusion is based on the proof of the
operator relation (Appendix~A)
\begin{displaymath}
\exp(-\,\Omega u
D^{1-\varepsilon})\,D^\alpha\,\exp(\Omega\,u\,D^{1-\varepsilon})=
(D^\varepsilon+\Omega)^{\alpha/\varepsilon},\quad
0<\varepsilon\leq 1,\quad\alpha\leq\varepsilon,
\end{displaymath}
where $u$ is a variable, and $\Omega$ a constant (all the
notations follow \cite{3}). However, the commutator (A2) in
\cite{3} does not hold true for any continuous functions. In
particular, by direct calculations one can found
\begin{displaymath}
[D^\alpha, \Omega
tD^{1-\varepsilon}]\,{t^{\alpha-1}}{\Gamma(\alpha)}=\Omega\,
(\alpha+\varepsilon-1)\,\frac{t^{\varepsilon-1}}
{\Gamma(\varepsilon)}
\end{displaymath}
rather than $\alpha\,\Omega\,t^{\varepsilon-1}/\Gamma
(\varepsilon)$ that it follows from \cite{3}. The cause lies in
vanishing the term $D^\alpha\,t^{\alpha-1}=0$. On the other hand,
if in time domain one expands the Havriliak-Negami (Cole-Cole)
response function as an infinite power series, its first term is
just $t^{\alpha-1}/\Gamma(\alpha)$ (in notation of \cite{3}). Thus
the expression (A5) is invalid for the HN response. The commutator
(A2) remains true only for the case, when $\varepsilon=1$ (CD
relaxation). The expression (A5) for the Cole-Davidson case can be
derived directly by the Leibniz's formula for fractional
derivatives without the operator identity.

In this connection it should also be recalled about difficulties
with the interpretation of fractional operators in terms of the
fractal Cantor set. Following \cite{3,21}, the memory function is
represented by a Cantor fractal function. However, such a memory
function possesses only a power-like property asymptotically, and
the approach itself requires else some average procedure for the
log-periodic term appearing together with the power \cite{23}. The
formalism \cite{21} is exactly macroscopic, but not stochastic.
The main feature of the complex system evolution lies just in a
stochastic background of dynamical processes. In the framework of
the approach \cite{21}, to find the relationship between the local
random characteristics of complex systems and the universal
deterministic empirical laws is hardly possible.

\section{Concluding Remarks}
In this paper we have shown the outlook of the probabilistic
approach proposed to the analysis of relaxation phenomena in the
complex systems. The approach permits ones to consider the
observable relaxation law on the unique theoretical base, the
limits theorem of probability theory. The general probabilistic
formalism treats the relaxation of the complex systems regardless
of the precise nature of local interactions. In a natural way, it
gives an efficient method for calculating the dynamical evaluating
averages of the relaxation processes. We have obtained all (known
up to now) the empirical relaxation laws, characterized their
parameters, connected the parameters with local random
characteristics of the relaxation processes, reconstructed the
internal structure of relaxing systems, justified the energy
criterion, demonstrated the transition from the analysis of the
microscopic random dynamics in the systems to the macroscopic
deterministic description by integro-differential equations. As a
rule, the classical methods of statistical physics take into
account the Central Limit Theorem in respect to the probability
distributions having finite invariance. However, the assumption
does not help to clarify the nature of relaxation phenomena. The
above approach has the advantage over the traditional models and
goes behind the classical statistical physics. The preliminary
results presented in this paper are promising and give a
confidence for a fundamental understanding of the relaxation
processes in the framework.

\section*{Acknowledgements}
The author acknowledges Prof. Karina Weron for useful discussions
on the subject.


\end{document}